\documentclass[usegraphicx]{mn2e}
\usepackage{txfonts}
\def\beq{\begin{equation}}
\def\eeq{\end{equation}}
\def\kms{\,\rm km\,{s}^{-1}}

\def\mpc{\,\rm Mpc}
\def\LCDM{\Lambda{\rm CDM}}

\title[Structures in galaxy cluster fields]{Complex structures in galaxy cluster fields: implications for gravitational lensing mass models }

\author[L. J. King and V. L. Corless]{
Lindsay King$^{*}$ and Virginia Corless$^{*}$\\
$^{*}$Institute of Astronomy, Madingley Rd, Cambridge CB3 0HA, UK}

\date{
Accepted ........
Received .......;
in original form ......}
\pubyear{2006}
\begin{document}
\maketitle
\begin{abstract}
The distribution of mass on galaxy cluster scales is an important test of 
structure formation scenarios, providing constraints on the nature of 
dark matter itself. Several techniques have been used to probe the mass 
distributions of clusters, sometimes yielding results which are discrepant, or 
at odds with clusters formed in simulations - for example
giving NFW concentration parameters much higher than expected in the 
standard CDM model. In addition, the velocity fields of some well studied 
galaxy clusters reveal the presence of several structures close to the line-of-sight, 
often not dynamically bound to the cluster itself.
We investigate what impact such neighbouring but unbound massive structures 
would have on the determination of cluster profiles using weak gravitational lensing. 
Depending on its concentration and mass ratio to the primary halo, one secondary 
halo close to the line-of-sight can cause the estimated NFW concentration parameter to be 
significantly higher than that of the primary halo, and also cause the estimated 
mass to be biased high. Although it is difficult to envisage how this mechanism alone 
could yield concentrations as high as reported for some clusters, multiple haloes close to 
the line-of-sight, such as in the case of Abell 1689, can substantially increase the 
concentration parameter estimate. Together with the fact that clusters are triaxial, and 
that including baryonic physics also leads to an increase in the concentration of a dark 
matter halo, the tension between observations and the standard CDM model is eased. 
Additionally, we note that if the alignment with the secondary structure is imprecise, then 
the estimated concentration parameter can also be even lower than that of the primary 
halo, reinforcing the importance of identifying structures in cluster fields.  
\end{abstract}

\def\A{{\cal A}}
\def\eck#1{\left\lbrack #1 \right\rbrack}
\def\eckk#1{\bigl[ #1 \bigr]}
\def\rund#1{\left( #1 \right)}
\def\abs#1{\left\vert #1 \right\vert}
\def\wave#1{\left\lbrace #1 \right\rbrace}
\def\ave#1{\left\langle #1 \right\rangle}
\def\arcsecf {\hbox{$.\!\!^{\prime\prime}$}}
\def\arcminf {\hbox{$.\!\!^{\prime}$}}
\def\bet#1{\left\vert #1 \right\vert}
\def\vp{\varphi}
\def\vt{{\vartheta}}
\def\d{{\rm d}}
\def\eps{{\epsilon}}
\def\vc{\vec} 
\def\s{{\rm d}}
\def\s{{\rm s}}
\def\t{{\rm t}}
\def\E{{\rm E}}
\def\L{{\cal L}}
\def\q{{\rm \i}}


\section[]{Introduction}
Galaxy clusters are the most massive bound structures in the universe, 
sign-posting dark matter and acting as laboratories inside which the nature 
of dark matter can be investigated (e.g. Spergel \& Steinhardt 2000). In addition, the cluster mass function is a sensitive test of cosmological parameters (e.g. Bahcall et al. 2002).
Several techniques have been used to estimate cluster profiles and masses, including X-ray studies, dynamics of cluster members and gravitational lensing. For clusters to be useful as cosmological
probes, remaining systematics in these methods must be addressed. 

One such systematic is the presence of large-scale structure along the line-of-sight. Hoekstra (2003) has estimated that uncorrelated large-scale structure can increase errors on weak lensing determined virial masses by a factor of two. A means to minimize the impact of large-scale structure in the context of cluster detection has been proposed by Maturi et al. (2005). 
Another concern is non-linear structure close to the line-of-sight: for example, Lokas et al. (2006) performed an analysis of the velocities of galaxies in the Abell 1689 field, a $\sim10^{15} M_{\odot}$ cluster at $z=0.18$, concluding that dynamical estimates of the cluster mass are extremely sensitive to which galaxies are considered to be cluster members or to belong to other structures. They suggest that unbound structures, clearly seen in velocity space, might also impact on the lensing analysis of this cluster. It is this issue that we address here, in the broader context of massive cluster lenses. 

Abell 1689 is not atypical in complexity, both considering other cluster fields (e.g. Cl0024+1654, Czoske et al. 2002) and from simulations. As an illustration of the latter, we searched the Virgo-Millennium Database (Springel et al. 2005; Lemson et al. 2006) for haloes more massive than $10^{15} M_{\odot}$ (the mass being taken within a radius where the halo has an overdensity 200 times the critical density of the simulation)  in the $z\sim 0.2$ output of the simulation. Of the 13 haloes contained in that volume (a cube of 500$h^{-1}\mpc$ on a side),  3 have at least 1 halo of mass ratio less than 20:1 within a comoving radius of 8$h^{-1}\mpc$.

A model which is commonly used to describe virialized haloes is the NFW profile (Navarro, Frenk \& White 1996; 1997; hereafter NFW), which is found to be a good fit to roughly 70\% of clusters in simulations (Jing 2000). The NFW model is often parametrized with a mass $M_{200}$ (roughly equivalent to a virial mass) and a concentration parameter $c$. Values of $c$ at odds with the distribution obtained from
cosmological simulations for structures of similar mass have recently 
been claimed: for example, from a combined weak and strong lensing analysis of Abell 1689, Broadhurst et al. (2005) determined $c=14\pm 1.5$, typical of {\em galaxy} mass haloes, whereas $c\sim 4$ is expected for massive cluster haloes (e.g. NFW).
The fact that clusters are triaxial and typically not 
spherically symmetric goes some way towards explaining the apparently high concentration 
parameters. Using N-body simulations, Clowe, De Lucia \& King (2004) demonstrated that halo triaxiality causes a scatter in the estimates of $c$ and $M_{200}$. Oguri et al. (2005) found that about 6\% of cluster scale haloes can reproduce the profile of Abell 1689 at the $2-\sigma$ level, with the long axis very close to the line-of-sight. They note that consistency with their Abell 1689 data falls to $<0.1\%$ of clusters when spherically symmetric models are assumed as in Broadhurst et al. (2005), even taking the $3-\sigma$ limits of the observations. In other clusters, triaxiality may more easily account for the value derived from observations when spherical symmetry has been assumed (e.g. MS 2137-23; Gavazzi 2005).

We estimate what impact physically close structures can have on cluster mass profiles determined in a weak lensing analysis, using simple NFW models for the primary cluster and for the additional haloes. In the next section we introduce the main principles behind the simulations described in section 3. Our results are summarised in section 4, and we discuss our findings in section 5.

\section[]{Weak Lensing Background}

We focus on weak lensing, potentially by multiple haloes close to the line-of-sight.
For a review of weak lensing, see Bartelmann \& Schneider (2001), and for a summary of multiple lens plane theory see for example Schneider, Ehlers \& Falco (1992).

We consider $N$ haloes each with an associated lens plane, at which deflection of light from a distant source occurs. The lens plane corresponding to the halo closest to the observer is the image plane, 
the observer's sky. The Jacobian matrix of the lens mapping ${\cal A}$ encapsulates the distortion induced
by a gravitational lens; after $N$ deflections ${\cal A}_{\,\rm tot}$ is given by
\begin{equation}
{\cal A} _{\,\rm tot} = {\cal I} - \sum_{i=1}^{N}{\cal U} _i {\cal
A}_i\;,
\end{equation}
where $\cal I$ is the identity matrix and 
\begin{equation}
{\cal U}_i = \left( \begin{array}{cc}
	\kappa^{(i)}+\gamma_{1}^{(i)}  & \gamma_{2}^{(i)} \\
	\gamma_{2}^{(i)}  & \kappa^{(i)}-\gamma_{1}^{(i)}\\
	\end{array}
	\right)
\end{equation}
for the $i$th deflection. The convergence $\kappa$ (dimensionless surface mass density) causes an isotropic magnification and the (complex) shear $\gamma$ introduces a distortion. For the $j$th lens, the intermediate Jacobian matrix is
\begin{equation}
{\cal A}_j = {\cal I} - \sum_{i=1}^{j-1}\beta_{ij}{\cal U}_i{\cal A}_i\;;~~~~~ {\cal A}_1={\cal I}
\end{equation}
where
\begin{equation}
\beta  _{ij} = \frac{D_{\rm os}}{D_{{\rm o}j}}\frac{D_{ij}}{D_{i{\rm s}}},
\end{equation}
with $D_{xy}$ being the angular diameter distances between $x$ and $y$, where o, s, $i$ and $j$ denote the observer, source, $i$th lens and $j$th lens respectively.

For example, 3 lens planes would result in a Jacobian matrix
\begin{eqnarray}
{\cal A}&=& {\cal I} - {\cal U}_1-{\cal U}_2-{\cal U}_3+\beta_{12}{\cal U}_1{\cal U}_2+\beta_{13}{\cal U}_1{\cal U}_3\\\nonumber&+&\beta_{23}{\cal U}_2{\cal U}_3
-\beta_{12}\beta_{23}{\cal U}_1{\cal U}_2{\cal U}_3\;.
\end{eqnarray}

For structures physically close to the main cluster lens $\beta_{ij}$ is small (this is quantified below), so in that case we immediately drop terms which are of higher order than linear in $\beta_{ij}$. Furthermore, since we are concerned with weak lensing where $\kappa\ll1$ and $\gamma\ll1$, only the linear terms in ${\cal U}$ need be retained. Thus the Jacobian matrix in (1) simplifies to 
\begin{equation}
{\cal A}_{\rm tot}={\cal I}-\sum_{i=1}^{N}{\cal U}_{i}\,,
\end{equation}
the determinant of which yields the magnification
\begin{equation}
\mu^{-1}=\left(1-\sum_{i=1}^{N}\kappa^{(i)}\right)^2-\left(\sum_{i=1}^{N}\gamma^{(i)}_{1}\right)^2-\left(\sum_{i=1}^{N}\gamma^{(i)}_{2}\right)^2\,,
\end{equation}
and where the effective shear and convergence are simply
\begin{equation}
\gamma=\left(\sum_{i=1}^{N}\gamma_{1}^{(i)},\sum_{i=1}^{N}\gamma_{2}^{(i)}\right)\,,~~~~\kappa=\sum_{i=1}^{N}\kappa^{(i)}\,.
\end{equation}

The shape and orientation of a weakly lensed galaxy projected on the sky is written as a complex 
ellipticity $\epsilon$, with modulus $|\epsilon|=[1-(b/a)]/[1+(b/a)]$ where $b/a$ is the axis ratio, and with a phase which is twice the position angle. This apparent ellipticity is related to the source ellipticity 
$\epsilon^{\s}$ and reduced shear, $g=\gamma/1-\kappa$, by
\begin{equation}
\epsilon=\frac{\epsilon^s+g}{1+g^*\epsilon^{\s}}\,,~~~~~ \approx\epsilon^s+\gamma\;,
\label{sheq}
\end{equation}  
where * denotes complex conjugation.

The lensed and unlensed ellipticity probability distributions are related through
\begin{equation}
p_{\epsilon}=p_{\epsilon^{\s}}\left
|\frac{{\rm d^{2}\epsilon^{\s}}}{{\rm d^{2}\epsilon}}\right |\;,
\label{ninfo}
\end{equation}
and the expectation value for the lensed ellipticity on a patch of sky $\ave\epsilon = g \approx \gamma$  (e.g. Schramm \& Kayser 1995). This is the basis of using the shapes of distant galaxies to constrain
lens models.

The number counts of galaxies are also changed: although a lens magnifies a source, an element of area is also changed by lensing. A nice description of this competition between effects can be found in Canizares (1982). This magnification effect results in local lensed number counts $n$ being related to the unlensed counts $n_{\rm o}$ and to the slope of these counts $\alpha$ by  $n=n_{\rm o}\mu^{\alpha-1}$.

A particular family of parametrized lens models is described by a set of parameters $\Pi$ ($c$ and $M_{200}$ in the case of an NFW model). The values of these parameters which best fit the
observed $n_{\gamma}$ lensed galaxy ellipticities (galaxy $i$ at position $\vc\vt_{i}$ having ellipticity $\epsilon_{i}$) are obtained by minimizing the shear log-likelihood function (Schneider, King \& Erben 2000; King \& Schneider 2001; King, Schneider \& Springel 2002):
\begin{equation}
\ell_{\gamma}=-\sum_{i=1}^{n_{\gamma}}{\rm\ln\;}p_{\epsilon}(\epsilon_{i}|g(\vc\vt_{i});\Pi)\;.
\label{liksh}
\end{equation}  
                     
\section{Simulations}
The simulations involve lensing populations of galaxies at $z=1$ using various composite lenses, comprising a cluster at $z=0.2$ and nearby structures. The best-fitting parameters for {\em single} lenses which would give rise to the simulated catalogues are obtained by application of a maximum likelihood technique, minimizing (11). 

The background source redshift is typical of the mean of source galaxies in weak lensing studies - the actual redshift {\em distribution} becomes important only for lenses at $z\gtrsim0.3$ (e.g. Seitz \& Schneider 1997) when the lensing effectiveness is more sensitive to source redshift. Galaxies are randomly distributed on the sky over a 30\arcmin field, with Gaussian distributed intrinsic ellipticities having dispersion $\sigma_{\epsilon^s}$. A number density of 30 galaxies/arcmin$^{2}$ is used, characteristic of ground-based observations, and Poisson noise is accounted for. The slope of the number counts is taken to be $\alpha=0.5$ for the purposes of incorporating source depletion due to magnification. 

Each component of the composite lens is modeled using an NFW halo; the convergence and shear for this are given by e.g. Bartelmann (1996). Our fiducial primary cluster has $c_{\rm p}=5$, and 
throughout the total mass is fixed at $M_{\rm tot}=10^{15} M_{\odot}$. The values of NFW concentration parameters for the primary $c_{\rm p}$ and secondary structures $c_{\rm s}$ are consistent with profiles of haloes formed in simulations (e.g. NFW; Eke, Navarro \& Steinmetz 2001). We assume a  $\LCDM$ cosmology with matter density parameter $\Omega_{\rm m}=0.3$, cosmological constant $\Omega_{\Lambda}=0.7$, and a Hubble constant $H_0=70\kms\mpc^{-1}$. 

\begin{figure}
\begin{center}
\includegraphics[width=3in]{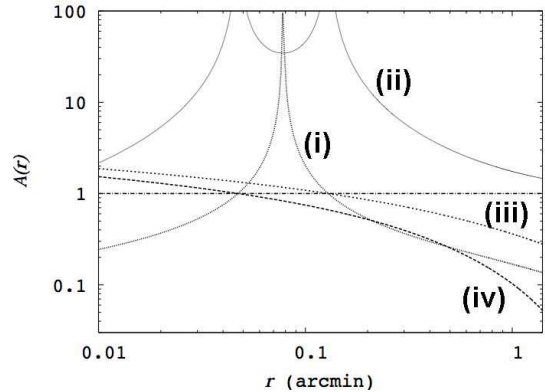}
\caption{The magnitude $A(r)$ of (i) the reduced shear $g$, (ii) magnification $\mu$, (iii) mean convergence $\bar\kappa$ (=1 at tangential critical curve),  for our fiducial NFW lens ($c=5$, $M_{200}=10^{15}$ M$_{\odot}$). The function $2\kappa-\bar\kappa$ is also plotted (iv); this gives the location of the radial critical curve. The lens is at $z=0.2$ and sources are at $z=1$.}
\label{qs}
\end{center}
\end{figure}

Although we focus on weak lensing, the strong lensing region towards the centre of our fiducial lens sheds light on the behaviour of various quantities of importance. As a function of distance from the centre $r$, Fig.\,\ref{qs} shows $\left|g\right|$, $\left|\mu\right|$, $\bar\kappa$ (mean convergence inside $r$) and $2\kappa-\bar\kappa$.
For a circularly symmetric profile, the solutions of {\bf det}${\cal A}=0$ 
(where $\mu$ formally diverges) show that the tangential critical 
curve occurs when $\bar\kappa(r)=1$ and the radial critical curve when $2\kappa-\bar\kappa =1$. 
The Einstein radius is strongly dependent on $c$ for a given virial mass:
for our 10$^{15}M_{\odot}$ halo, when $c=5$, 
$\theta_{E}\approx 8''$ and when $c=10$, $\theta_{E}\approx 27''$. 

For a secondary structure at z=0.22 (0.24), from (4) the parameter $\beta\approx 0.1 (0.2)$; coupled with the fact that we work in the weak lensing regime this confirms our choice to use (6)-(10) from above. To retain generality it is assumed that the various structures are at the same redshift; the impact of offset redshifts is considered separately at the end of the next section. 
\section{Results}
\subsection{Single or precisely aligned haloes}
\begin{figure}
\begin{center}
\includegraphics[width=2.9in]{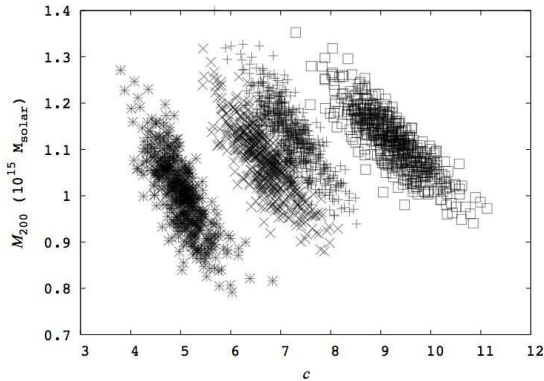}
\caption{The concentration parameter $c$ and mass $M_{\,200}$ for best-fitting single NFW haloes to (i) single fiducial halo (*), (ii) 10\% of mass in secondary halo, $c_{\rm s}=7$ (x), (iii) same as (ii) but with 20\% in secondary halo (+), (iv) 10\% of mass in each of two $c_{\rm s}=10$ haloes ($\square$). Each point within a group corresponds to a different noise realisation of random unlensed galaxy positions and ellipticities.}
\label{simuls}
\end{center}
\end{figure}

For each lens discussed in this subsection, 500 catalogues of lensed galaxies (with random positions and ellipticities, $\sigma_{\epsilon}^{\rm s}=0.3$) were generated, and best-fitting parameters recovered through the maximum likelihood method. The data were taken from the weak lensing regime, estimated in practice either from the shear or the location of giant arcs.
If all the mass, $10^{15}M_{\odot}$, is contained within the primary cluster $c_{\rm p}=5$, there is a scatter in recovered parameters as illustrated in Fig.\,\ref{simuls}. If space-based observations were available, the scatter would be decreased due to the (factor of a few) increase in the number of sources useful for a lensing analysis (see King \& Schneider (2001) for treatment of individual 
NFW haloes). 

Next consider one secondary halo directly aligned along the line of 
sight, having 10\% of the total mass, and with $c_{\rm s}=7$. Fig.\,\ref{simuls} 
shows the best-fitting $M_{200}$ and $c$ when a single component NFW model is 
fit to the lensing data. The value of $M_{200}$ is quite close to the $total$ mass 
in the system - of course over-estimating the mass of the primary cluster since lensing 
is sensitive to the {\em projected} mass. Of particular note is that the best-fitting
$c$ is much larger than $c_{\rm p}$. Increasing the mass of the 
secondary structure so that it contains 20\% of the total mass exacerbates the trend 
towards high values of $c$. Similar, more extreme, effects are found when we consider $c_{\rm s}=10$. 
For precisely aligned haloes, a less massive secondary structure (with a higher 
concentration parameter) biases the mass estimate of the cluster (or even of the system as a whole), and more importantly results in a higher estimate of $c$. For a composite lens the best-fitting single NFW model parameters are those with a reduced shear most consistent with the probability distribution 
of lensed ellipticities. Similar to the case of a single lens, noise 
results in a scatter, and there is a clear degeneracy direction along which 
models have very similar $g$, as seen in Fig.\,\ref{simuls}. So the key to 
understanding why a structure which contains a small fraction of the mass of 
the system should drive $c$ to a larger or smaller value is simply the 
behaviour of $g(r)$ over the radial range considered for analysis.

In Fig.\,\ref{simuls} we also show the best-fitting models when two haloes, each 
containing 10\% of the mass, and each with $c_{\rm s}$=10 are along the line-of-sight to the primary 
halo. It is also interesting to divide the total mass $M_{\rm tot}$ into $N$ equal 
aligned haloes, in which case we find numerically that the best-fitting single NFW halo 
concentration parameter scales as $c_{\rm s}\,N^{2/5}$, whereas the mass scales as 
$M_{\rm tot}\,N^{1/5}$.   
\subsection{Misaligned haloes}
\begin{figure}
\begin{center}
\includegraphics[width=2.9in]{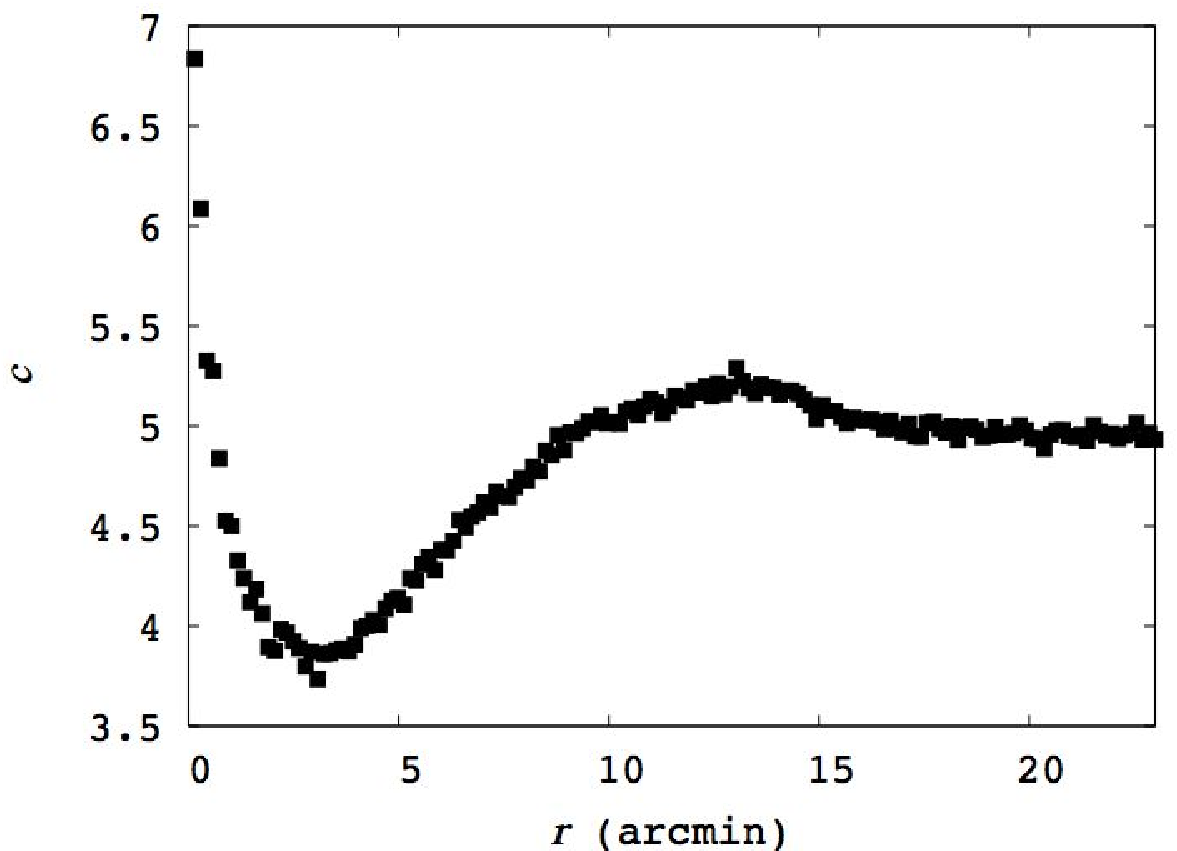}
\includegraphics[width=2.9in]{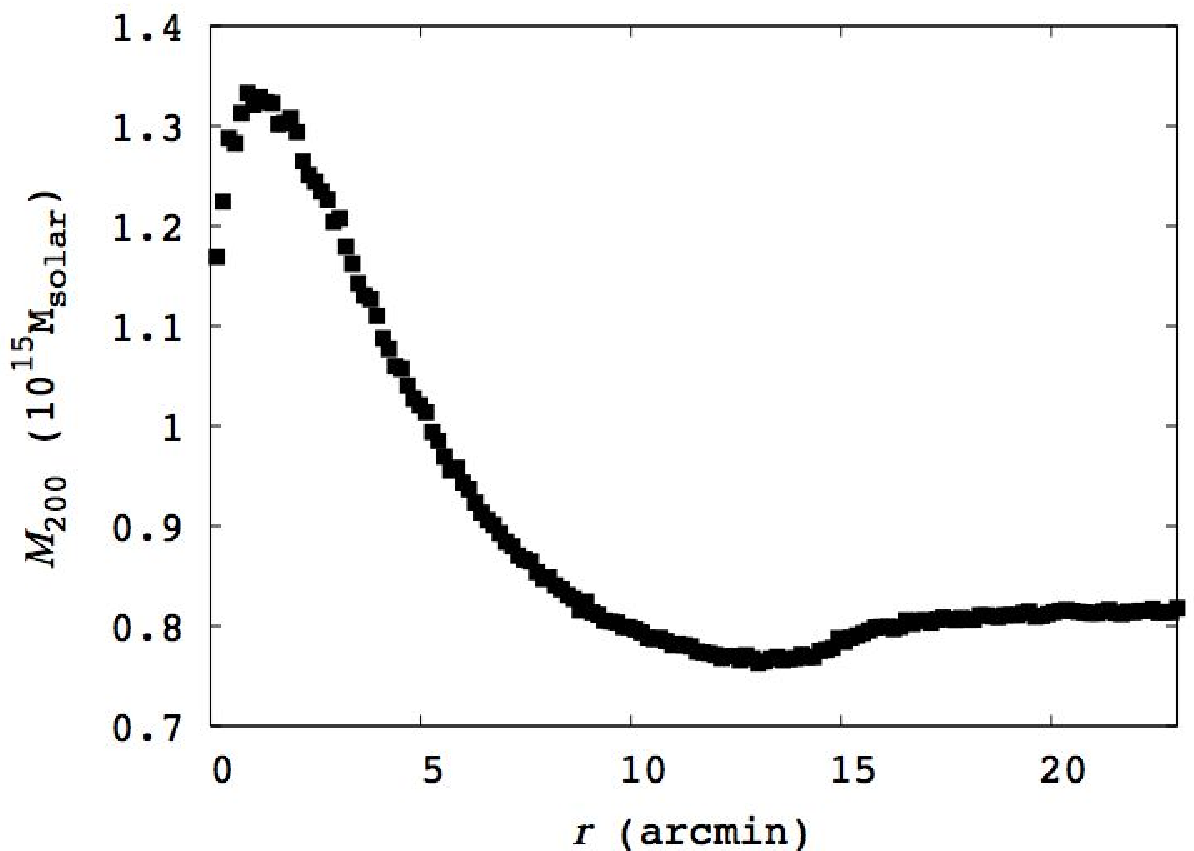}
\includegraphics[width=2.9in]{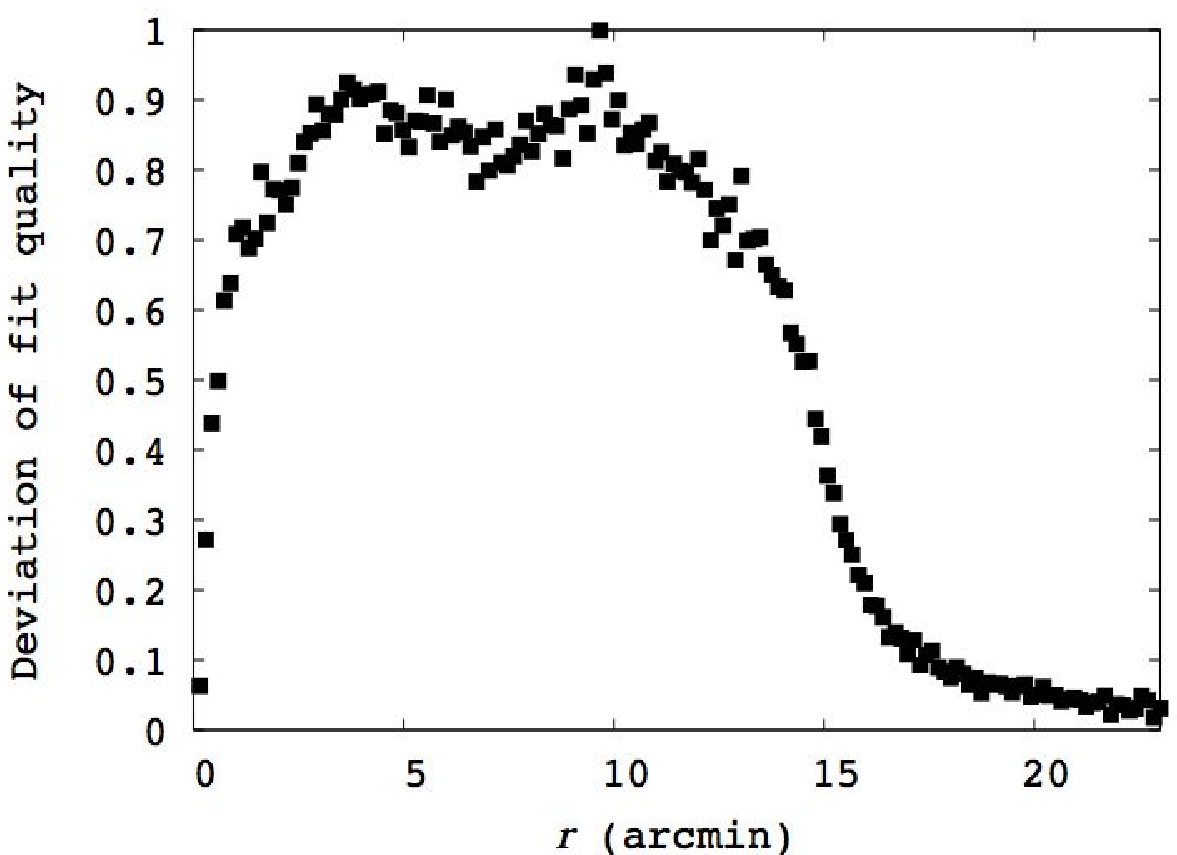}
\caption{The behaviour of a single NFW fit concentration parameter $c$ (upper panel) and mass $M_{\,200}$ (middle panel) to the weak lensing data from a composite lens (see text for details) as a function of projected component separation $r$. For each data point, the positions and ellipticities of the unlensed background galaxy population are fixed. The quality of fit is indicated in the bottom panel: 0 
is the fit achieved with the primary component in isolation and 1 is the poorest fit at intermediate separations.}
\label{unalign}
\end{center}
\end{figure}

We now explore the impact on the best-fitting single halo NFW parameters when the alignment between a primary halo and secondary halo is no longer precise. At some point, the presence of a massive secondary may be apparent due to strongly lensed arcs, but these may be difficult to identify 
given confusion with arcs produced by the primary (e.g. since substructure causes deviations from the regular tangential distortion associated with a smooth mass). For illustration, it is assumed that 20\% of the mass is contained in a secondary halo, $c_{\rm s}=7$. Fig.\,\ref{unalign} summarises the behaviour of the fit parameters as a function of separation; since we are interested in how separation changes the fit, the same set of random galaxy positions and a small ellipticity dispersion ($\sigma_{\epsilon}=0.05$) is used for each case. 

Moving the secondary further away from the line-of-sight to the primary first of all results in a reduction in $c$, when the misalignment is less than 
a couple of primary scale radii ($r_{\rm s}=r_{200}/c$). Beyond this, $c$ increases again, roughly 
until the secondary is completely outside $r_{200}$ of the primary. When the secondary 
leaves the field, $c$ tends to 5 as expected. The mass estimate also tends to that of the primary, 0.8$\times 10^{15} M_{\odot}$ at this point. Again, over a range of component separations, the mass is both over- and under-estimated. The overall quality of fit is indicated in the bottom panel of Fig.\,\ref{unalign}; since the log-likelihood associated with a fit depends on factors such as the number density of galaxies, ellipticity dispersion and properties of the lenses, we use a scale such that the a fit to a single isolated halo gives a deviation in fit quality of 0, and the poorest fit case (at intermediate separations) gives a deviation of 1. 
\subsection{Dependence on halo redshift}
The halo redshift enters into the critical surface mass density for lensing, and into the 
critical density of the universe (through the Hubble parameter $H(z)$) at that epoch which relates $r_{200}$ to $M_{200}$. For aligned haloes, taking $z_{\rm s}=0.24$ and $c_{\rm s}=7$ with 20\% of the total mass results in less than a few \% shift in $c$ and in $M$, reinforcing the generality of our results for various lens configurations in redshift.

\section{Discussion and Conclusions}

In this paper we have briefly addressed the extent to which secondary haloes close to 
massive galaxy clusters impact on their weak lensing mass profiles. We use simple 
models for the cluster and for the secondary haloes, simulate lensing through these and determine the best-fitting single NFW halo parameters. We focus on low mass ratios, to estimate the maximum impact on the lensing analysis.

Non-linear structures close to the line-of-sight to a more massive cluster {\em do} have 
an impact on the estimated profile and mass of the cluster. It is difficult to attribute the very 
high reported values of $c$ to one secondary unbound halo, although this is more effective 
when $c_{\rm s}$ is high and the mass ratio is low. For example, Fig.\,\ref{unalign} shows that for a system of $10^{15} M_{\odot}$, with $c_{\rm p}=5$, $c_{\rm s}=7$ and 20\% of the mass contained in the secondary, the alignment has to be better than an arcminute to give an estimated $c$ which exceeds $c_{\rm p}$ by at least 10\%. More progress can however be made by 
having multiple haloes close to the line-of-sight. Besides over-estimating $c$ when the alignment is fairly precise, over a wide range of separations of a primary and secondary, $c$ can also be {\em under}-estimated, as shown in Fig.\,\ref{unalign}. Since a single NFW halo is fit to the data, the quality of the fits is better whenever there is close alignment, becomes poorer as the separation increases, improves again as the secondary starts to leave the data field, and returns to the same quality as the fit to a single halo whenever the secondary is well outside the data field.

A point which will be addressed in further work is the probability distribution for {\em observed} values of $c$ from our models. Perhaps the observed distribution of lenses is skewed towards those with components in close alignment to the line-of-sight (with a higher surface mass density and higher lensing efficiency) - thus though $c$ may be biased high only over a small range of projected separations, as noted in the previous paragraph, there would be a bias towards these lenses. Also, since $c$ is likely to be under-estimated when there is an additional structure outside the central regions of the primary halo, there is a greater chance that extra strong lensing features due to it would be recognised, or that a mass reconstruction would at some point detect it. We have tested the efficiency of the latter, and it strongly depends on the mass ratio and separation of the structures, as well as galaxy number density. Additionally, the fit of a single halo would be poor in that regime (see Fig.\,\ref{unalign}) as noted in the previous paragraph.

It has been shown that triaxiality can go some way towards accounting for high 
values of $c$ (Gavazzi et al. 2005; Oguri et al. 2005), and in a forthcoming 
paper we explore the properties of this class of model in detail (Corless \& King in prep.). 
Another consideration is that real clusters are typically compared with those 
formed in dark matter simulations - neglecting the impact of baryons. Cosmological simulations including gas dynamics, star formation and 
radiative cooling show that dark matter halo profiles are expected to become more 
concentrated as gas cools in their inner regions (e.g. Gnedin et al. 2004).
To triaxiality and baryonic physics we add unbound structures in close alignment with the
main halo as a possible contributor to observed inconsistency with expectations. 
A general caveat: in interpreting departure from NFW in individual clusters as a sign of failure of the CDM paradigm, it should be remembered that many well studied cluster lenses are among the 30\% of cases referred to by Jing (2002) which are not relaxed, isolated systems - and therefore not 
necessarily expected to be well described by an NFW profile, even after baryonic physics has 
been accounted for.

In the context of Abell 1689, it is consistent with the available lensing and spectroscopic data (see for example the compilation in Lokas et al. 2006 and references therein) that the field consists of a main triaxial halo whose long axis is close to the line-of-sight (as suggested by Oguri et al. 2005) and other haloes close to the line-of-sight to it (which might also be similarly oriented). The positions of the galaxies in the structures at different velocities overlap on the plane of the sky, indicating that they are along the line-of-sight (Lokas et al. 2006). From the currently available spectroscopic data, it is difficult to say how precise the alignment is between these structures. However, Andersson \& Madejski (2004) noted the coincidence in X-ray observations between a high-redshift gas region approximately 1$\arcmin$ offset from the main cluster and a grouping of optically identified high redshift giant ellipticals. Since the Einstein radius of Abell 1689 is larger than that of our fiducial cluster, this degree of alignment would be in the regime which would result in an over-estimate of $c$. 
More detailed dynamical modeling and estimation of the masses of the structures in the Abell 1689 field will be possible using new spectroscopy (in excess of 500 cluster members) taken by Czoske (2004). In general, such samples with large numbers of galaxies and well defined selection criteria will prove very useful in building up a 3-D picture 
of galaxy clusters, when combined with gravitational lensing and X-ray studies.

\section{acknowledgments}
This work was supported by the Royal Society (LJK) and by the Marshall Foundation (VLC). 
We thank Vincent Eke, Martin Haehnelt, Antony Lewis, Max Pettini and Mark Wilkinson for helpful discussions and the referee for constructive comments. The Millennium Simulation databases used in this paper and the web application providing online access to them were constructed as part of the activities of the German Astrophysical Virtual Observatory.

\def\ref#1{\bibitem[1998]{}#1}

\end{document}